\documentclass[prc,twocolumn,showpacs,10pt,floatfix]{revtex4}

\usepackage{graphicx}
\usepackage{amsmath}
\usepackage{amssymb}
\usepackage{citesort}

\newcommand{\identity}[1]{#1}  

\DeclareRobustCommand{\captionopensquare}{\resizebox{0.7em}{!}{$\square$}}

\DeclareRobustCommand{\captionopencircle}{\raisebox{-0.3ex}{\resizebox{0.9em}{!}{$\circ$}}}

\DeclareRobustCommand{\captionopenuptriangle}{\resizebox{0.8em}{!}{$\vartriangle$}}

\DeclareRobustCommand{\captionsoliddiamond}{\resizebox{0.8em}{!}{\rotatebox{45}{$\blacksquare$}}}
\DeclareRobustCommand{\captionopendiamond}{\resizebox{0.8em}{!}{\rotatebox{45}{$\square$}}}

\newcommand{\hbarfactor}{\ensuremath{\frac{\hbar^2}{2B}}}

\newcommand{\hbarinvfactor}{\ensuremath{\frac{2B}{\hbar^2}}}
\newcommand{\hbarinvfactorinline}{\ensuremath{2B/\hbar^2}}
\newcommand{\Ai}{\ensuremath{\mathrm{Ai}}}  
\newcommand{\basistag}{\ensuremath{\alpha=0}}

\begin{document}


\title{Consequences of wall stiffness for a $\beta$-soft potential}

\author{M. A. Caprio}
\affiliation{Center for Theoretical Physics, Sloane Physics Laboratory, 
Yale University, New Haven, Connecticut 06520-8120, USA}

\date{\today}

\begin{abstract}
Modifications of the infinite square well E(5) and X(5) descriptions
of transitional nuclear structure are considered.  The eigenproblem
for a potential with linear sloped walls is solved.  The consequences
of the introduction of sloped walls and of a quadratic transition
operator are investigated.
\end{abstract}

\pacs{21.60.Ev, 21.10.Re, 27.70.+q}

\maketitle


\section{Introduction}
\label{secintro}

The E(5) and X(5) models have been proposed by
Iachello~\cite{iachello2000:e5,iachello2001:x5} to describe the
essential characteristics of shape-transitional forms of quadrupole
collective structure in nuclei.  The E(5) model, for $\gamma$-soft
nuclei, and the X(5) model, for axially symmetric nuclei, are both
based upon the approximation of the potential energy as a square well
in the Bohr deformation variable $\beta$.  These models produce
predictions for level energy spacings and electromagnetic transition
strengths intermediate between those for spherical oscillator
structure and for deformed
$\gamma$-soft~\cite{wilets1956:oscillations} or deformed
axially-symmetric rotor~\cite{bohr1998:v2} structures.

\begin{figure}
\begin{center}
\includegraphics[width=0.9\hsize]{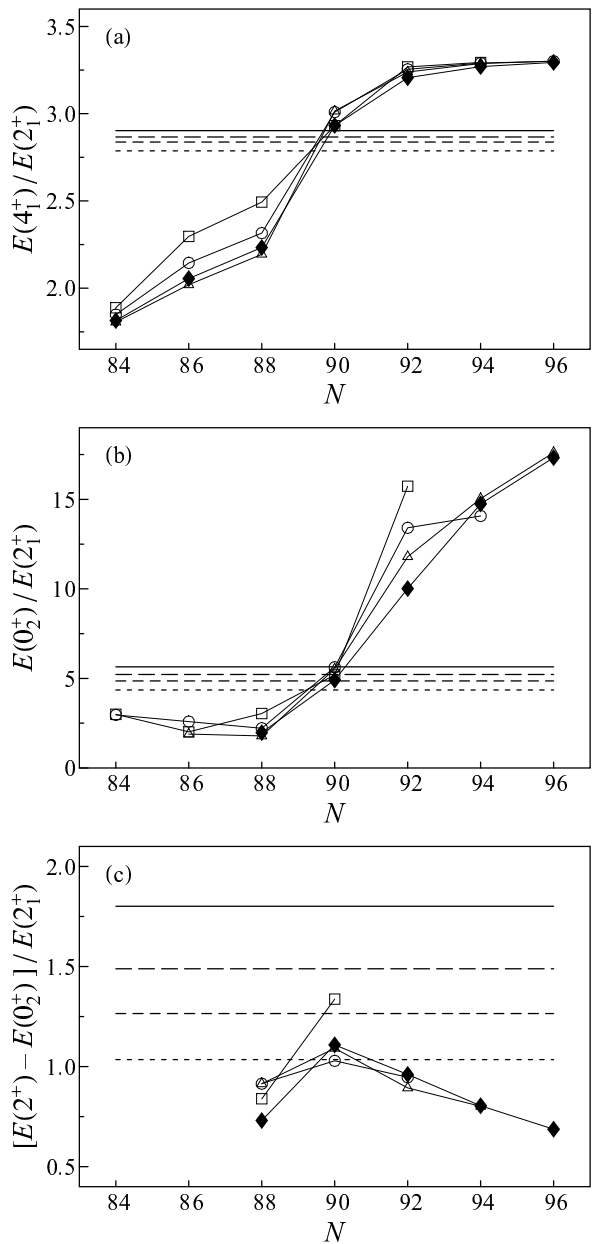}
\end{center}
\vspace{-12pt}
\caption
{Evolution of the (a) $4^+_1$ energy, (b) $0^+_2$ energy, and (c)
energy spacing scale of the excited $0^+$ sequence, all normalized to
the $2^+_1$ energy, across the $N$=90 transition region, for the
Nd~(\captionopensquare), Sm~(\captionopencircle),
Gd~(\captionopenuptriangle), and Dy~(\captionsoliddiamond) isotopic
chains.  Shown for comparison are the X(5) predictions (solid line)
and the present sloped well predictions for various values of the
parameter $S$ defined in~(\ref{eqnsdef})~--- $S$$=$100 (long-dashed
line), $S$$=$50 (short-dashed line), and $S$$=$25 (dotted line).  Data
are from Refs.~\cite{nds1995:150,nds1996:152,nds1998:154,ndsboth:156}.
(Figure based upon Ref.~\cite{caprio2002:156dy-beta}.)
\label{figsystchains}
}
\end{figure}

The X(5) predictions for level energy spacings and electromagnetic
transition strengths have been extensively compared with data for
nuclei in transitional regions between spherical and rotor
structure~\cite{casten2001:152sm-x5,bizzeti2002:104mo-x5,kruecken2002:150nd-rdm,brenner2002:x5-a80a100,caprio2002:156dy-beta,hutter2003:104mo106mo-rdm,clark2003:x5-search,mccutchan2004:162yb-beta,bijker2003:x5-gamma}.
For several such nuclei, including the $N$=90 isotopes of Nd, Sm, Gd,
and Dy, the X(5) predictions match well the yrast band level energies
and the excitation energy of the $K^\pi$$=$$0^+_2$
band head~[Fig.~\ref{figsystchains}(a,b)].  The X(5) predictions
also reproduce essential features of the electric quadrupole
transitions from the $K^\pi$$=$$0^+_2$ band to the ground state band:
the presence of strong spin-ascending interband transitions but
highly-suppressed spin-descending transitions.%
%
%

However, several discrepancies exist between the X(5) predictions and
observed values.  The spacing of level energies in the
$K^\pi$$=$$0^+_2$ band is predicted to be much larger than in the
ground state band, but empirically at most a slightly larger energy
scale is found for the $K^\pi$$=$$0^+_2$ band~[Fig.~\ref{figsystchains}(c)]~\cite{caprio2002:156dy-beta,clark2003:x5-search,mccutchan2004:162yb-beta}.
This overprediction is encountered in descriptions of transitional
nuclei with the interacting boson model~(IBM) and geometric collective
model~(GCM) as well~\cite{scholten1978:ibm-fits,zhang1999:152sm-gcm}.
For nuclei with yrast band level energies matching the X(5)
predictions, the yrast band $B(E2)$ strengths tend to fall below the
X(5) predictions, and sometimes even below the pure rotor predictions
(see Fig.~2 of Ref.~\cite{clark2003:x5-search}).  For the $N$=90
nuclei, the transitions between the $K^\pi$$=$$0^+_2$ and ground state
bands have strength ratios typically matching those predicted, but
their strength scale is considerably weaker than
predicted~\cite{casten2001:152sm-x5,kruecken2002:150nd-rdm,caprio2002:156dy-beta,clark2003:n90-x5,casten2003:n90-x5-comment}.%

It is thus necessary to ascertain which aspects of the X(5)
description are most important in determining the predictions for
these basic observables.  The square well potential involves an
infinitely-steep ``wall'' in the potential as a function of $\beta$,
presumably a radical approximation.  Moreover, the model has so far
been used only with a first-order electric quadrupole transition
operator, but the likely importance of second-order effects has been
noted by Arias~\cite{arias2001:134ba-e5} and by Pietralla and
Gorbachenko~\cite{pietrallaXXXX:beta-soft}.  In the present work, the
infinitely stiff confining wall is replaced with a gentler, sloped
wall, constructed using a linear potential.  The effects upon
calculated observables of the introduction of a sloped wall and of a
quadratic transition operator are addressed.  A computer code for
solution of the sloped well eigenproblem is provided through the
Electronic Physics Auxiliary Publication Service~\cite{swell-epaps}.

\section{Solution method}

Consider the Bohr Hamiltonian~\cite{bohr1998:v2}
\begin{multline}
\label{eqnbohrh}
H=- \frac{\hbar^2}{2B} \Biggl[ \frac{1}{\beta^4}
\frac{\partial}{\partial \beta}
\beta^4  \frac{\partial}{\partial \beta} + \frac{1}{\beta^2 \sin 3\gamma} 
\frac{\partial}{\partial \gamma} \sin 3\gamma \frac{\partial}{\partial \gamma} 
\\
 - \frac{1}{4 \beta^2} \sum_\kappa \frac{M_\kappa^2}{\sin^2(\gamma -
\frac{2}{3} \pi \kappa)} \Biggr]
+V(\beta,\gamma),
\end{multline}
where $\beta$ and $\gamma$ are the Bohr deformation variables and the
$M_\kappa$ are angular momentum operators, with potential
\begin{equation}
\label{eqnpotential} V(\beta)=\begin{cases} 0 & \beta\leq \beta_w\\ C
(\beta-\beta_w) & \beta>\beta_w .
\end{cases}
\end{equation}
Since this potential is a function of $\beta$ only, the
five-dimensional analogue of the central force problem arises.  The
usual separation of ``radial'' ($\beta$) and ``angular''
variables~\cite{wilets1956:oscillations,rakavy1957:gsoft} occurs,
yielding eigenfunctions of the form
$\Psi(\beta,\gamma,\omega)$$=$$f(\beta)\Phi(\gamma,\omega)$, where
$\omega$$\equiv$$(\vartheta_1,\vartheta_2,\vartheta_3)$ are the Euler
angles.  The angular wave functions $\Phi(\gamma,\omega)$, common to
all $\gamma$-independent problems, are known~\cite{bes1959:gamma}.
For the radial problem, following Rakavy~\cite{rakavy1957:gsoft}, it
is most convenient to work with the ``auxiliary'' radial wave function
$\varphi(\beta)$$\equiv$$\beta^2f(\beta)$.  This function obeys a
one-dimensional Schr\"odinger equation with a ``centrifugal'' term,
\begin{equation}
\label{eqnradial}
\left[-\hbarfactor\frac{\partial^2}{\partial\beta^2}
+\hbarfactor\frac{\alpha}{\beta^2}+V(\beta)-E\right]\varphi(\beta)=0,
\end{equation}
where the centrifugal coefficient $\alpha$ is related to the O(5)
separation constant $\tau$ ($\tau$$=$$0,1,\ldots$) by
$\alpha$$=$$(\tau+1)(\tau+2)$.  For problems with a more general
potential $V(\beta,\gamma)$$=$$V_\beta(\beta)+V_\gamma(\gamma)$,
Iachello~\cite{iachello2001:x5} showed that an approximate separation
of variables occurs, provided that $V_\gamma(\gamma)$ confines the
nucleus to $\gamma$$\approx$$0$ (see Ref.~\cite{iachello2001:x5} for
details).  In this ``$\gamma$-stabilized'' case, the eigenfunctions
are of the form
$\Psi(\beta,\gamma,\omega)$$\propto$$f(\beta)\eta(\gamma)\phi_{KLM}(\omega)$,
where the $\phi_{KLM}(\omega)$ are the conventional rigid rotor
angular wave functions~\cite{bohr1998:v2} for angular momentum $L$,
$z$-axis projection $M$, and symmetry axis projection $K$.  The
auxiliary radial wave function again obeys~(\ref{eqnradial}), but now
with $\alpha$$=$$\frac{1}{3}L(L+1)+2$.  

In the region $\beta$$<$$\beta_w$, the potential $V(\beta)$
of~(\ref{eqnpotential}) vanishes, and the radial
equation~(\ref{eqnradial}) reduces to the Bessel equation of order
$\nu$$=$$(\alpha+1/4)^{1/2}$.  The solutions with the correct
convergence properties at the origin are
$\varphi(\beta)$$\propto$$\beta^{1/2}J_\nu(\varepsilon^{1/2}\beta)$, where
$\varepsilon$$\equiv$$(\hbarinvfactorinline) E$.  In the region
$\beta$$>$$\beta_w$, where the potential is linear in $\beta$, an
analytic solution does not exist for the full problem with centrifugal
term.  For $\alpha$$=$$0$ only, (\ref{eqnradial}) reduces to the Airy
equation, with solutions
$\varphi(\beta)$$\propto$$\Ai[c^{1/3}(\beta-\beta_w)-c^{-2/3}\varepsilon]$,
where $c$$\equiv$$(\hbarinvfactorinline) C$.

\begin{figure}
\begin{center}
\includegraphics[width=1.0\hsize]{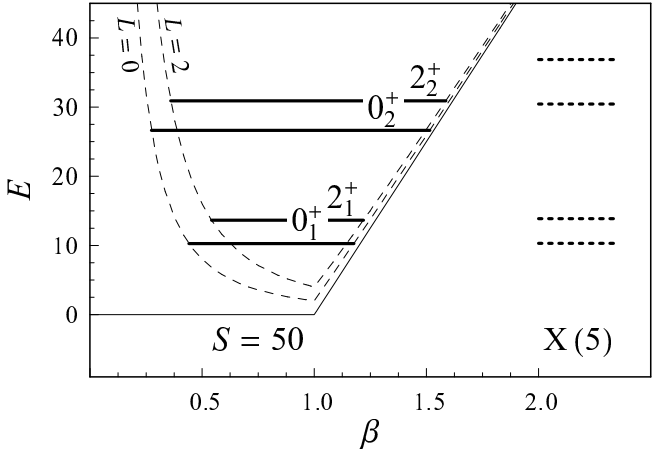}
\end{center}
\vspace{-12pt}
\caption
{Energies of low-lying $0^+$ and $2^+$ levels for the sloped well
potential with $S$=50.  The potential without the five-dimensional
centrifugal term is shown (solid curve), together with the potential
including the centrifugal contributions for $L$$=$$0$ and $L$$=$$2$
(dashed curves).  The energies of the corresponding states for the
X(5) model are shown for comparison at right.  [The $S$$=$$50$
calculation is for $\hbarinvfactorinline$$=$$1$ and $\beta_w$$=$$1$,
while the X(5) calculation is scaled to $\beta_w$$=$$1.40$ to provide
the same ground state eigenvalue.]
\label{figcentrif}
}
\end{figure}

The analytic solutions obtained for $\alpha$$=$$0$ provide a very
efficient basis for numerical diagonalization to obtain the true
$\alpha$$\neq$$0$ solutions of the radial equation~(\ref{eqnradial}).
It is first necessary to obtain a basis set of $\alpha$$=$$0$
solutions
\begin{equation}
\varphi_i^{\basistag}(\beta)=\begin{cases} N_1
\beta^{1/2}J_{1/2}[(\varepsilon_i^{\basistag})^{1/2}\beta] &
\beta\leq \beta_w\\ 
N_2 \Ai[c^{1/3}(\beta-\beta_w)-c^{-2/3}\varepsilon_i^{\basistag}] &
\beta>\beta_w .  \end{cases}
\end{equation}
The eigenvalues of $\varepsilon$ are determined by the condition that
$\varphi(\beta)$ be continuous and smooth at the matching point
$\beta$$=$$\beta_w$.  This yields a transcendental equation which is
solved numerically for $\varepsilon$.  The normalization coefficients
$N_1$ and $N_2$ then follow from continuity and the requirement
$\int_0^\infty d\beta |\varphi(\beta)|^2$$=$$1$.  Since the radial
equation~(\ref{eqnradial}) has the form of a one-dimensional
Schr\"odinger equation, its solution for general values of $\alpha$
may be carried out as the matrix diagonalization problem for a
corresponding ``Hamiltonian'' matrix $h$, including the
centrifugal potential, with respect to these $\alpha$$=$$0$ basis
functions, with entries
\begin{equation}
\label{eqnradialhij}
h_{ij}\equiv\delta_{i,j}\varepsilon_i^{\basistag}+\alpha\int_0^\infty
d\beta \, \varphi_i^{\basistag}(\beta)
\, \frac{1}{\beta^2}\,  \varphi_j^{\basistag}(\beta).
\end{equation}
Convergence in this basis is rapid --- for instance, the eigenvalues
of the ground state and first excited radial solution converge
to within $\sim$$1.5\%$ of their true values with a truncated basis of
only $5$ eigenfunctions.  Values shown in this paper are calculated
for a basis size of 25.  For illustration, an example potential, with
centrifugal contribution, and the corresponding calculated eigenvalues
are shown in Fig.~\ref{figcentrif}.%

Electromagnetic transition strengths can be calculated from the matrix
elements of the collective multipole operators.  The general $E2$
operator for the geometric
model~\cite{gneuss1971:gcm,hess1981:gcm-pt-os-w,eisenberg1987:v1} may be expanded in
laboratory frame coordinates $\alpha_{2\mu}$ as~\cite{endnote-e2conjugate}
\begin{equation}
\label{eqnme2lab}
\mathfrak{M}(E2;\mu)=A_1\alpha_{2\mu}
+A_2[\alpha\times\alpha]^{(2)}_\mu+\cdots .
\end{equation}  
For the present purposes, it is necessary to reexpress this operator
in terms of the intrinsic frame coordinates and
$D^2(\omega)$~\cite{endnote-dfunction}, giving, to second order in
$\beta$,
\begin{multline}
\label{eqnme2intrinsic}
\mathfrak{M}(E2;\mu)=A_1\beta\bigl[D^{2\,*}_{\mu0}\cos\gamma+\frac{1}{\sqrt{2}}(D^{2\,*}_{\mu2}+D^{2\,*}_{\mu-2})\sin\gamma\bigr]\\
-\sqrt{\frac{2}{7}}A_2\beta^2\bigl[D^{2\,*}_{\mu0}\cos2\gamma-\frac{1}{\sqrt{2}}(D^{2\,*}_{\mu2}+D^{2\,*}_{\mu-2})\sin2\gamma\bigr].
\end{multline}

In both the $\gamma$-independent and $\gamma$-stabilized cases, the
matrix element of $\mathfrak{M}(E2;\mu)$ between two eigenstates
factors into an angular integral and a radial integral.  Here we
consider matrix elements between unsymmetrized $\gamma$-stabilized
wave functions~\cite{bohr1998:v2}
\begin{equation}
\Psi_{\alpha K J M}=\left(\frac{2J+1}{8\pi^2}\right)^{1/2}
D^{J\,*}_{MK}(\omega)\,\Phi_{\alpha KJ}(\beta,\gamma),
\end{equation}
as needed in calculations for the rigid rotor, X(5), or 
$\gamma$-stabilized sloped well models.  The matrix element separates
into intrinsic and Euler angle integrals, yielding
\begin{multline}
\left\langle\Psi_{\alpha' K' J' M'} ||
\mathfrak{M}(E2;\mu)||\Psi_{\alpha K J M} \right\rangle = \\
(-)^{J'-J}(J'K'2(K-K')|JK)\left[A_1
I_1-\sqrt{\frac{2}{7}}A_2I_2\right],
\label{eqndoublebarme}
\end{multline}
in terms of
\begin{equation}
\begin{aligned}
I_1&\equiv\int d\tau \,\Phi_{\alpha' K'J'}^*(\beta,\gamma)\,
\beta\cos\gamma \,
\Phi_{\alpha KJ}(\beta,\gamma)
\\
I_2&\equiv\int d\tau \,\Phi_{\alpha' K'J'}^*(\beta,\gamma) \,
\beta^2\cos2\gamma \,
\Phi_{\alpha KJ}(\beta,\gamma)
\end{aligned}
\end{equation}
for $K'$$-$$K$$=$$0$ or
\begin{equation}
\label{eqnI1I2deltak2}
\begin{aligned}
I_1&\equiv\int d\tau \,\Phi_{\alpha' K'J'}^*(\beta,\gamma)\,
\frac{1}{\sqrt{2}} \beta\sin\gamma \,
\Phi_{\alpha KJ}(\beta,\gamma)\\
I_2&\equiv\int d\tau \,\Phi_{\alpha' K'J'}^*(\beta,\gamma)\, ( - )
\frac{1}{\sqrt{2}}\beta^2\sin2\gamma \,
\Phi_{\alpha KJ}(\beta,\gamma),
\end{aligned}
\end{equation}
for $K'$$-$$K$$=$$\pm2$, where $d\tau$$\equiv$$\beta^4d\beta \left|
\sin 3\gamma \right| d\gamma$ and the reduced matrix element
normalization convention is that of Rose~\cite{rose1957:am}.  (The matrix elements of the
\textit{symmetrized} wave functions, for $K$$\neq$$0$, may be
calculated from this matrix element as usual~\cite{bohr1998:v2}.)
Considering the present $\beta$-$\gamma$ separated wave functions
$\Phi(\beta,\gamma)$$=$$f(\beta)\eta(\gamma)$, for the case of no
$\gamma$ excitation (so $K'$$=$$K$$=$$0$),
and under the
approximation $\gamma$$\approx$$0$, these integrals reduce to
$I_1$$=$$\int\beta^4d\beta f_{\alpha'K'J'}(\beta) \beta f_{\alpha
KJ}(\beta)$ and $I_2$$=$$\int\beta^4d\beta f_{\alpha'K'J'}(\beta) \beta^2
f_{\alpha KJ}(\beta)$.  Transition strengths are $B(E2;J$$\rightarrow$$
J')$$=$$\linebreak[0](2J'+1)\linebreak[0]/(2J+1)\linebreak[0]\,|\left<J'||
\mathfrak{M}(E2)||J\right>|^2$.  Quadrupole moments, defined by
$eQ_{J}$$\equiv$$(16\pi/5)^{1/2}\linebreak[0]\,\langle J J | \mathfrak{M}(E2;0) | J J
\rangle$, may be calculated as
$eQ_{J}$$=$$(16\pi/5)^{1/2}\linebreak[0](JJ20|JJ)\linebreak[0]\,\left<J||
\mathfrak{M}(E2)||J\right>$.

The following calculations can be considerably simplified if it is
noted that the eigenvalue spectrum and wave functions depend upon the
Hamiltonian parameters $B$, $\beta_w$, and $C$ only in the combination
\begin{equation}
\label{eqnsdef}
S\equiv \hbarinvfactor \beta_w^3 C,
\end{equation}
to within an overall normalization factor on the eigenvalues and
overall dilation of all wave functions with respect to $\beta$.  [This
follows from invariance of the Schr\"odinger equation solutions under
multiplication of the Hamiltonian by a constant factor and under a
transformation of the potential
$V'(\beta)$$=$$a^2V(a\beta)$~\cite{caprio2003:gcm}.]  For a given
value of $S$, the numerical solution need only be obtained once, at
some ``reference'' choice of parameters (\textit{e.g.},
\hbarinvfactorinline$=$$1$ and $\beta_w$$=$1), and the solution for
any other well of the same $S$ can be deduced analytically.
Specifically, suppose the reference calculation yields an eigenvalue
$\varepsilon$ and a normalized radial wave function $f(\beta)$.  Then
a calculation performed for the same $B$ and $S$ but for a different
width $\beta_w'$ produces the eigenvalue $\varepsilon'$ and normalized
wave function $f'(\beta)$ given by the simple rescalings
\begin{equation}
\begin{gathered}
\label{eqnfwellscale}
\varepsilon'=\beta_w'^{-2}\varepsilon\\
f'(\beta)=\beta_w'^{-5/2} f(\beta/\beta_w'),
\end{gathered}
\end{equation}
and the radial integrals scale to $I_1'=\beta_w'I_1$ and
$I_2'=\beta_w'^2I_2$.  Thus, the essential parameter which controls
the relative strengths of the linear and quadratic terms of the $E2$
operator is $A'$$\equiv$$A_2\beta_w/A_1$, in terms of which the matrix
element in~(\ref{eqndoublebarme}) is
\begin{multline}
(-)^{J'-J}(J'K'2(K-K')|JK)\\
\times
A_1\beta_w\left[
I_1|_{\beta_w=1}-\sqrt{\frac{2}{7}}A'I_2|_{\beta_w=1}\right].
\label{eqndoublebarmescaled}
\end{multline}
Ratios of $E2$ matrix elements depend only upon $S$ and $A'$.

A computer code for solution of the sloped well eigenproblem and for
calculation of the radial matrix elements between eigenstates is
provided through the Electronic Physics Auxiliary Publication
Service~\cite{swell-epaps}.  This code also calculates observables for
the E(5) and X(5) models.

\section{Results}

In the following discussion, let us restrict our attention to
$\gamma$-stabilized structure relatively close to the X(5) limit of
the sloped well model, since this regime is most directly relevant to the
transitional nuclei recently considered in the context of the X(5)
model.  The sloped well potential approaches a pure linear potential
as $\beta_w$ vanishes at fixed slope (that is, as $S$$\rightarrow$$0$)
and approaches a square well as the slope goes to infinity at fixed
$\beta_w$ (that is, as $S$$\rightarrow$$\infty$).  It can thus produce
a much wider variety of structures than are considered in the present discussion.
However, calculations for the full range of these cases may be obtained with
the provided computer code~\cite{swell-epaps}.

First we examine the energy spectrum, comparing it to the X(5)
spectrum.  Naturally, the eigenvalues for the sloped well are lowered
relative to those for the X(5) well of the same $\beta_w$, as the
outward slope of the wall effectively widens the well, causing level
energies to ``settle'' lower.  The essential feature is that the
widening of the well introduced by the wall slope is a relatively
small fraction of the well width at low energies, while it is much
greater at high energies, as may be seen by inspection of the
potential (Fig.~\ref{figcentrif}).  Thus, the high-lying levels
experience a disproportionately greater increase in the accessible
range of $\beta$-values than do low-lying levels and consequently are
lowered in energy relative to the low-lying levels.

\begin{figure}
\begin{center}
\includegraphics[width=1.0\hsize]{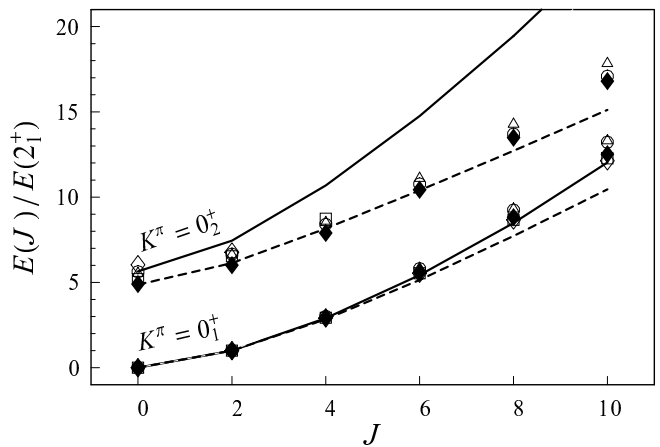}
\end{center}
\vspace{-12pt}
\caption
{Yrast and $K^\pi=0^+_2$ band level energies, normalized to
$E(2^+_1)$, for $^{150}$Nd~(\captionopensquare),
$^{152}$Sm~(\captionopencircle), $^{154}$Gd~(\captionopenuptriangle),
$^{156}$Dy~(\captionsoliddiamond), and
$^{162}$Yb~(\captionopendiamond).  The predictions for X(5) (solid curve) and
the sloped well with $S$$=$$50$ (dashed curve) are shown for
comparison.  Data are from
Refs.~\cite{nds1995:150,nds1996:152,nds1998:154,ndsboth:156,nds1999:162,mccutchan2004:162yb-beta}.
\label{figbandss50}
}
\end{figure}

From the calculated energies, it is seen that as $S$ is decreased from
infinity the higher-spin levels within a band are lowered more rapidly
than the lower-spin members, resulting in a reduction of the ratio
$R_{4/2}$$\equiv$$E(4^+_1)/E(2^+_1)$ for the yrast band
[Fig.~\ref{figsystchains}(a)] and a lowering of the curve of $E$
versus $J$ for each band (Fig.~\ref{figbandss50}).  The excited band
head energies are lowered as well [Fig.~\ref{figsystchains}(b)].  But
the most dramatic change is the rapid collapse of the spacing scale of
levels within the excited bands relative to that of the ground state
band [Figs.~\ref{figsystchains}(c) and~\ref{figbandss50}].  For $S$$\approx$$50$, the
predicted energy spacing scale within the $K^\pi$$=$$0^+_2$ band is
reduced sufficiently to be consistent with the spacings found for the
$N$$=$$90$ transitional nuclei, while the energies of low-spin yrast
band members and the $K^\pi$$=$$0^+_2$ band head are still relatively
close to their X(5) values, as shown in Fig.~\ref{figbandss50}.

\begin{figure*}
\begin{center}
\includegraphics[width=1.0\hsize]{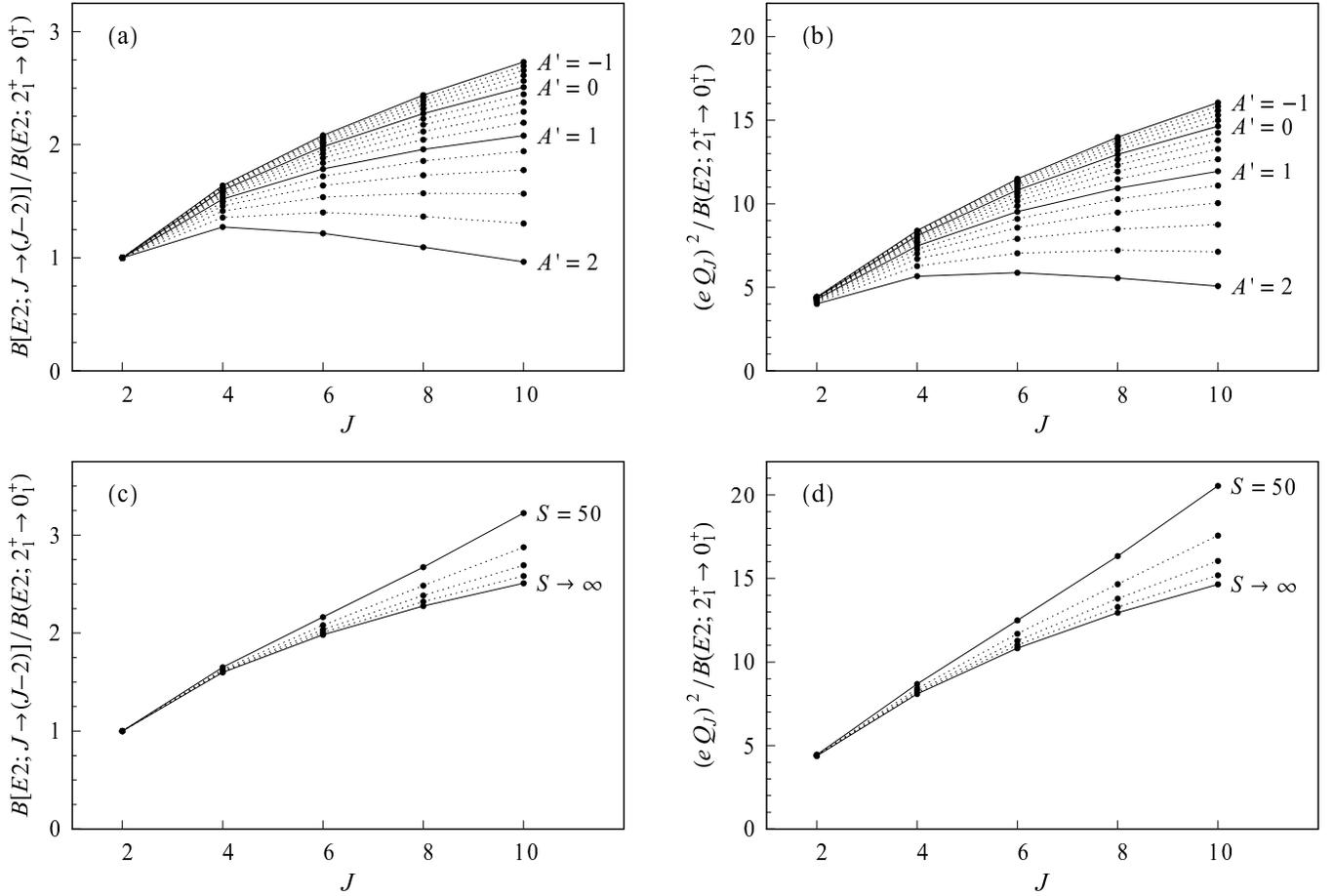}
\end{center}
\vspace{-12pt}
\caption
{Yrast band $B(E2)$ strengths and squared quadrupole moments,
normalized to $B(E2;2^+_1\rightarrow0^+_1)$.  (a,b)~Values for
the X(5) model ($S$$\rightarrow$$\infty$) calculated with a quadratic $E2$
transition operator, for $A'$ ranging from $-1$ to $2$ in equal steps.
(c,d)~Values for the sloped well ($S$$=$50, 100, 200, 500, and
$\infty$) calculated with a linear $E2$ transition operator.
\label{fige2curves}
}
\end{figure*}

\begin{figure*}
\begin{center}
\includegraphics[width=0.75\hsize]{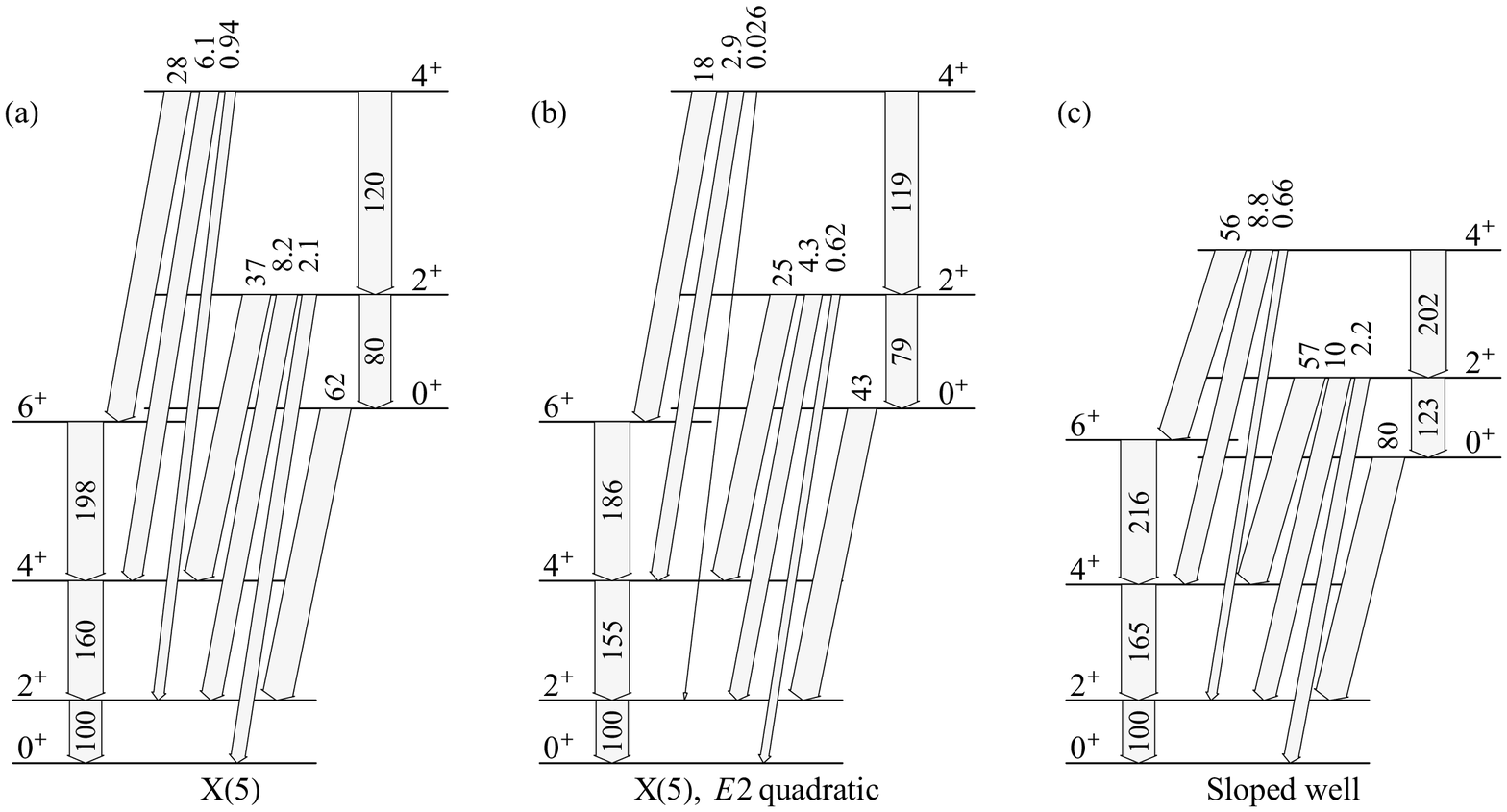}
\end{center}
\vspace{-12pt}
\caption
{Level schemes and selected $B(E2)$ strengths for (a)~the X(5) model
with a linear $E2$ operator, (b)~the X(5) model with a quadratic $E2$
operator ($A'$$=$$0.7$), and (c)~the sloped well ($S$$=$$50$) with a
linear $E2$ operator.  Arrow thicknesses
are proportional to the logarithm of the $B(E2)$ strength.
\label{figschemestheory}
}
\end{figure*}

The second order term in the $E2$ operator~(\ref{eqnme2intrinsic}) can
interfere either constructively or destructively with the first order
term.  For all transitions between low-lying levels considered here,
the radial integrals $I_1$ and $I_2$ in~(\ref{eqndoublebarmescaled})
have the same sign.  Thus, negative values of $A'$ lead to
constructive interference [note the negative coefficient
in~(\ref{eqndoublebarmescaled})], while positive values lead to
destructive interference.  For the X(5) square well, the higher-spin
members of the yrast band have larger average $\beta$ values than do
the low-spin members, so the quadratic term is relatively more
important for the higher-spin levels.  In the case of destructive
interference, the curve showing the spin dependence of $B(E2)$ values,
normalized to $B(E2;2^+_1\rightarrow0^+_1)$, falls below that obtained
with the simple linear $E2$ operator, as seen in
Fig.~\ref{fige2curves}(a).  The broad range of such curves obtained
experimentally (see Ref.~\cite{clark2003:x5-search}) can be
qualitatively reproduced with different values of $A'$.  Destructive
interference also reduces the interband $B(E2)$ strengths and the
in-band $B(E2)$ strengths within the $K^\pi$$=$$0^+_2$ band, relative
to $B(E2;2^+_1\rightarrow0^+_1)$ [Fig.~\ref{figschemestheory}(b)],
ameliorating the overprediction of interband strengths in the X(5)
model.  The spin-descending interband transitions in the X(5) model
have highly-suppressed linear $E2$ matrix elements, so these
transitions are very sensitive to even a small quadratic contribution.
Values of $A'$ which give only moderate modifications to the other
transitions can give complete destructive interference for these
spin-descending transitions.  The spin dependence of quadrupole
moments within the yrast band is shown in
Fig.~\ref{fige2curves}(b).%
\begin{figure}
\begin{center}
\includegraphics[width=\hsize]{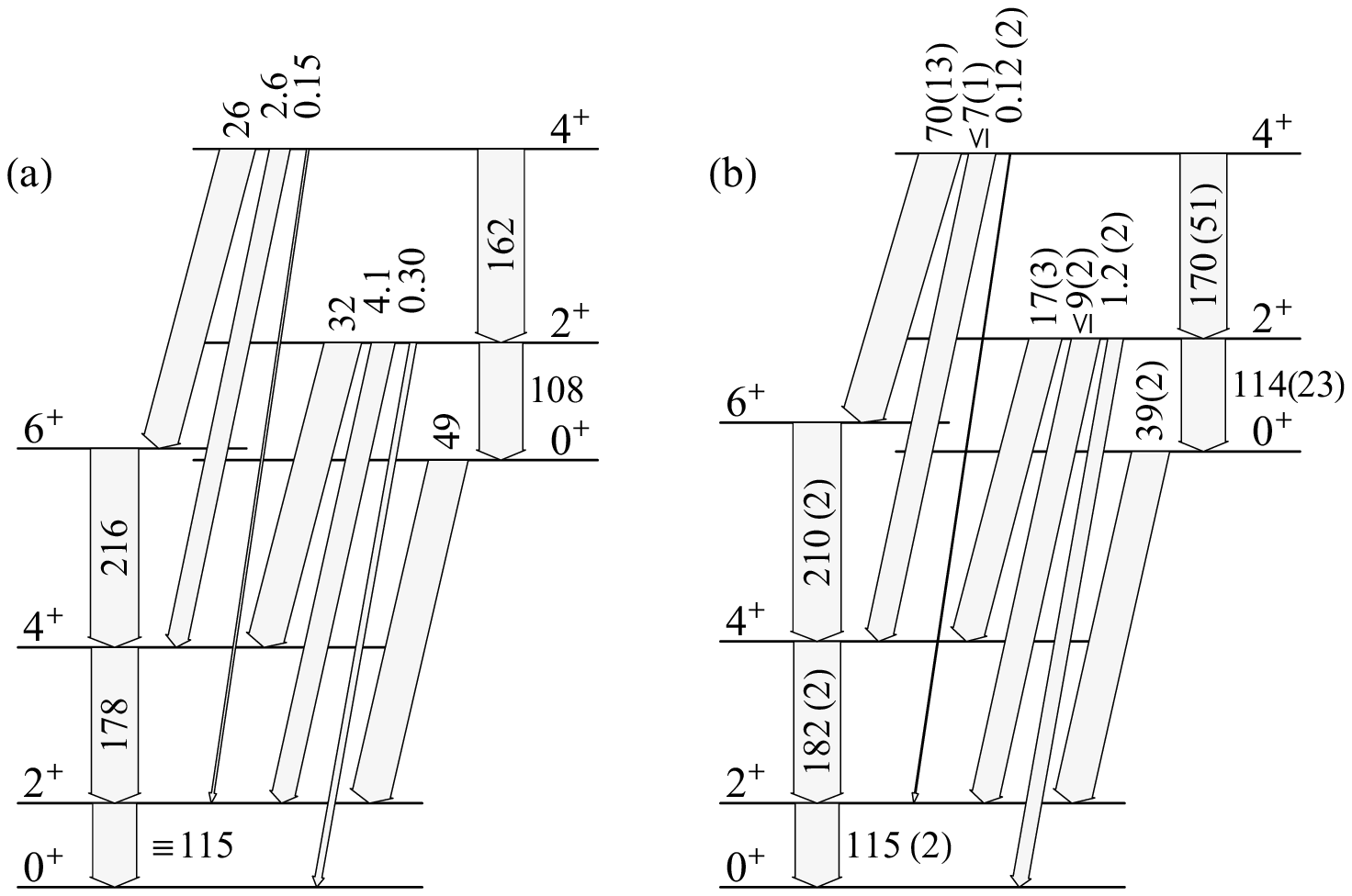}
\end{center}
\vspace{-12pt}
\caption
{Level scheme and selected $B(E2)$ strengths (a)~for the sloped well
with parameters chosen to approximately reproduce the observed
low-energy structure of $^{150}$Nd ($S$$=$$75$, $A'$$=$$0.6$) and
(b)~as measured for $^{150}$Nd~\cite{nds1995:150,kruecken2002:150nd-rdm}.  Arrow thicknesses
are proportional to the logarithm of the $B(E2)$ strength.  Limits are
indicated on experimental $B(E2)$ strengths for transitions with
unknown $E2/M1$ mixing ratios.
\label{figschemesexpt}
}
\end{figure}

Observe that the situation just described differs considerably from
that encountered for a pure rotor.  For a rigid rotor, the intrinsic
wave function $\Phi_{\alpha K}(\beta,\gamma)$ is the same for all
levels within a band, so $I_2$ provides only a uniform adjustment to
the intrinsic matrix element between bands.  Inclusion of the second
order term in $\mathfrak{M}(E2)$ thus leaves unchanged the ratio of any
two $B(E2)$ values within a band or the ratio of any two $B(E2)$ values between
the same two bands.

Although inclusion of a quadratic term in the $E2$ operator with
$A'$$>$$0$ can at least qualitatively explain the discrepancies
between the X(5) $B(E2)$ predictions and empirical values, this
explanation is not entirely satisfactory.  Many different
spin dependences of the $B(E2)$ values within the yrast band are
observed for nuclei with similar energy
spectra~\cite{clark2003:x5-search}, and these require correspondingly
varied, apparently \textit{ad hoc} choices of the parameter $A'$ for
their reproduction.  Moreover, it is possible to obtain estimates for
the coefficients $A_1$ and $A_2$ in the geometric $E2$ operator based
on a simple model of the nuclear charge and current distribution, as
described in Refs.~\cite{hess1981:gcm-pt-os-w,eisenberg1987:v1}, and
these values yield $A'$$\approx$$-0.2$, giving weak
\textit{constructive} interference for the low-lying transitions.  In
the interacting boson model, the $E2$ transition operator is of the
form
$T^{(E2)}$$\propto$$(d^\dag\times\tilde{s}+s^\dag\times\tilde{d})^{(2)}+\chi(d^\dag\times\tilde{d})^{(2)}$,
in terms of the boson creation operators $s^\dag$ and $d^\dag$, where
the value $\chi$$=$$-\sqrt{7}/2$ is commonly used in calculations
involving the transition from spherical to axially-symmetric deformed
structure~\cite{lipas1985:ecqf,iachello1987:ibm}.  In the classical
limit, $(d^\dag\times\tilde{s}+s^\dag\times\tilde{d})^{(2)}$ may be
approximately identified with the linear term of the geometric model
transition operator and $\chi(d^\dag\times\tilde{d})^{(2)}$ with the
quadratic term.  The addition of these terms is constructive for
low-lying transitions, and the relative contribution of the
second-order term is comparable to that obtained for $A'$$\approx$$-1$
in the present description.

The effect of sloped walls on the calculated $B(E2)$ strengths is
dominated by the greater broadening of the well at high energies than
at low energies discussed above.  While all the eigenfunctions
``spread'' in $\beta$ extent relative to those for the square well,
this spreading is most pronounced for the high-lying levels.  Since
the first order $E2$ operator is proportional to $\beta$, the $E2$
matrix elements tend to be enhanced for the higher-lying levels.  In
the yrast band, the in-band $B(E2)$ strengths for higher-spin band
members are increased relative to those for the lower-spin band
members, as are the quadrupole moments for higher-spin band members
[Fig.~\ref{fige2curves}(c,d)].  Several of the interband $B(E2)$
strengths are also increased relative to $B(E2;2^+_1\rightarrow0^+_1)$
(Fig.~\ref{figschemestheory}).  The changes in $B(E2)$ values induced
by decreasing $S$ are largely opposite in sense to those produced by
introduction of the second-order term in the $E2$ operator.  The
parameters $S$ and $A'$ may be chosen so as to balance these two
effects against each other, except that for the spin-descending
interband transitions the strong destructive interference tends to
dominate.

To allow comparison with empirical values, in Fig.~\ref{figschemesexpt}(a)
predictions obtained with the sloped wall potential and quadratic $E2$
operator are shown for parameter values chosen to approximately
reproduce the observed low-energy structure of $^{150}$Nd.  The
experimental values are given in Fig.~\ref{figschemesexpt}(b).

Finally, let us consider the effects of wall slope on the properties
of the $K^\pi$$=$$2^+_1$ band, or $\gamma$ band.  Within the
$\gamma$-stabilized separation of variables of
Ref.~\cite{iachello2001:x5}, the properties of this band are largely
independent of the specific choice of $\gamma$-confining potential
$V_\gamma(\gamma)$.  This potential determines the band head energy
as well as the $\gamma$-dependent wave function $\eta(\gamma)$.  The
wave function, however, simply contributes a normalization
factor $\int \left|\sin 3\gamma\right| d\gamma \eta_1(\gamma)\sin
\gamma \eta_0(\gamma)$ to $I_1$ in~(\ref{eqnI1I2deltak2}), and an
analogous factor to $I_2$, common to all electromagnetic matrix
elements between the $K^\pi$$=$$2^+_1$ band and the $K^\pi$$=$$0^+_1$
and $0^+_2$ bands.  Although these quantities can be calculated for
any particular hypothesized form for $V_\gamma(\gamma)$, such as a
harmonic oscillator potential~\cite{iachello2001:x5}, they in practice
may be treated as free parameters.

\begin{figure*}
\begin{center}
\includegraphics[width=\hsize]{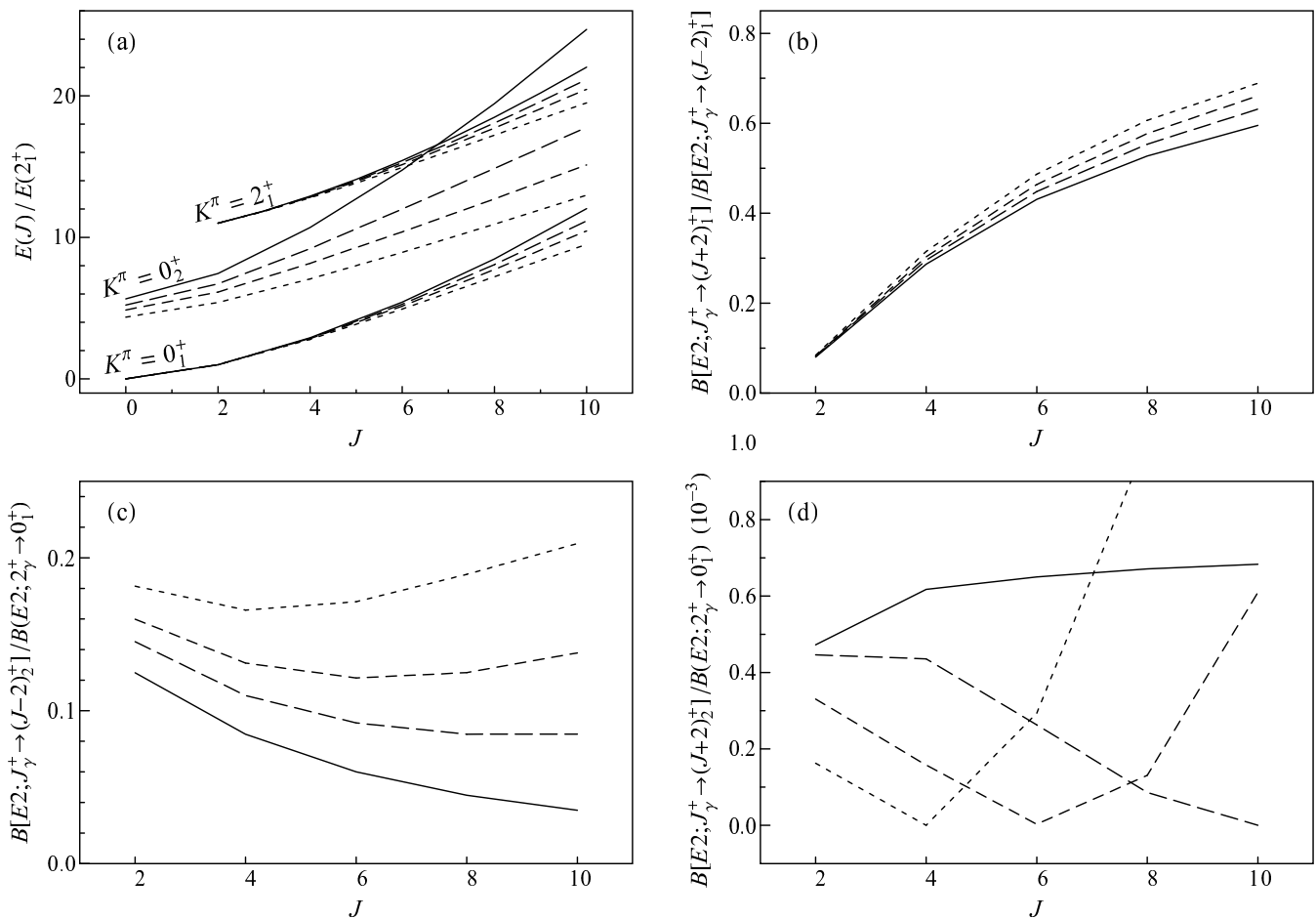}
\end{center}
\vspace{-12pt}
\caption
{Dependence of properties of the $K^\pi$$=$$2^+_1$ band upon wall
slope.  Values are shown for the X(5) model (solid line), $S$$=$100
(long-dashed line), $S$$=$50 (short-dashed line), and $S$$=$25 (dotted
line).  (a)~Yrast, $K^\pi$$=$$0^+_2$, and $K^\pi$$=$$2^+_1$ band energies,
illustrating the identical dependences of the yrast and
$K^\pi$$=$$2^+_1$ bands.  The $K^\pi$$=$$2^+_1$ band head energy is
arbitrary (see text).  (b)~$B(E2)$ branching ratios for $\Delta
J$$=$$\pm2$ transitions from the $K^\pi$$=$$2^+_1$ band to the yrast
band.  (c,d)~$B(E2)$ strengths for $\Delta J$$=$$\pm2$ transitions
from the $K^\pi$$=$$2^+_1$ band to the $K^\pi$$=$$0^+_2$ band,
normalized to $B(E2;2^+_\gamma\rightarrow0^+_1)$.  The separation of variables of
Refs.~\cite{iachello2001:x5,bizzeti2002:104mo-x5} has been used for all calculations.
\label{figgammaband}
}
\end{figure*}

The essential feature of the $K^\pi$$=$$2^+_1$ band is that the radial
wave function for each of its members is the ``ground state'' solution
of the radial equation~(\ref{eqnradial}) for the given angular
momentum.  This $K^\pi$$=$$2^+_1$ band is thus essentially a duplicate
of the yrast band, displaced to a higher energy by the excitation
energy in the $\gamma$ degree of freedom, with energy spacings and
radial wave functions for the even spin members identical to those for
the yrast band, but with the addition of odd spin members and with
different angular wave functions.  (Note that for $K$$\neq$$0$
Bijker~\textit{et al.}~\cite{bijker2003:x5-gamma} use a different
separation procedure from that in
Refs.~\cite{iachello2001:x5,bizzeti2002:104mo-x5}, yielding a modified
form of the radial equation with
$\alpha$$=$$\frac{1}{3}[L(L+1)-K^2]+2$, which changes the energy
spacings and in-band radial matrix elements by $\lesssim$$5\%$
relative to those of the yrast band.)  Thus, the dependence of
$K^\pi$$=$$2^+_1$ band properties upon wall slope closely matches that
of the yrast band properties.  Notably, the $K^\pi$$=$$2^+_1$
band does not demonstrate the rapid decrease in energy spacing scale
with decreasing wall slope exhibited by the $K^\pi$$=$$0^+_2$ band, as
illustrated in Fig.~\ref{figgammaband}(a).  This is at least qualitatively
consistent with the observed similarity of the yrast and
$K^\pi$$=$$2^+_1$, but not $K^\pi$$=$$0^+_2$, band energy
spacings in the $N$$=$90 X(5) candidate nuclei.

Since the even spin members of the $K^\pi$$=$$2^+_1$ band possess the
same radial wave functions as the yrast band members, the strengths of
transitions within the $K^\pi$$=$$2^+_1$ band or between this band and
the yrast band depend upon the same radial matrix elements as do the
yrast in-band transition strengths and quadrupole moments already
considered.  Consequently, decreasing wall slope leads to a moderate
enhancement of the interband transition strengths involving higher
spin levels, directly commensurate with the increases shown in
Fig.~\ref{fige2curves}(c,d).  The dependence of branching ratios from
the $K^\pi$$=$$2^+_1$ band to the yrast band on wall slope is shown in
Fig.~\ref{figgammaband}(b). Transitions between the $K^\pi$$=$$2^+_1$
band and the $K^\pi$$=$$0^+_2$ band depend instead upon radial matrix
elements which contribute to the $K^\pi$$=$$0^+_2$ to yrast band
transition strengths.  (The small radial matrix element values which
yield the characteristic suppression of spin-\textit{descending}
transitions from the $K^\pi$$=$$0^+_2$ band to the yrast band here
yield a suppression of spin-\textit{ascending} $K^\pi$$=$$2^+_1$ to
$0^+_2$ transitions.)  Decreasing wall slope yields enhancement of the
allowed transitions and, for the low-spin levels, either little change
or substantial reduction of the suppressed transitions
[Fig.~\ref{figgammaband}(c,d)].  Detailed quantitative predictions for
the $K^\pi$$=$$2^+_1$ band level energies and electromagnetic
observables, using either the separation of variables of
Refs.~\cite{iachello2001:x5,bizzeti2002:104mo-x5} or that of
Ref.~\cite{bijker2003:x5-gamma}, may be obtained with the provided
code~\cite{swell-epaps}.

\section{Conclusion}

The use of a $\beta$-soft potential within the geometric picture has
recently received  attention as providing a simple description
of nuclei intermediate between spherical and rigidly deformed
structure.  From the present results, it is seen that the energy
spacing scale of states within excited bands is highly sensitive to
the stiffness of the well boundary wall.  A potential for which the
well width increases with energy can produce a more compact spacing
scale for excited states than is obtained with a pure square well,
providing much closer agreement with the observed energy spectra for
nuclei in the $N$$\approx$$90$ transition region.  It is also found
that a second-order contribution to the $E2$ transition operator can
lead to a wide range of possible yrast band $B(E2)$ spin dependences,
as well as to modifications of off-yrast matrix elements.  However, a
systematic understanding of the proper strength for this second-order
contribution is needed if the $E2$ operator is to be applied
effectively.

\begin{acknowledgments}
Discussions with F.~Iachello, N.~V.~Zamfir, R.~F.~Casten,
N.~Pietralla, E.~A.~McCutchan, L.~Fortunato, and A.~Leviatan are
gratefully acknowledged.  This work was supported by the US DOE under
grant DE-FG02-91ER-40608 and was carried out in part at the European
Centre for Theoretical Studies in Nuclear Physics and Related Areas
(ECT*).
\end{acknowledgments}

\vfil



\begin{thebibliography}{36}
\bibitem{endnote-dfunction}
\bibinfo{name}{The $D$ function used here is that of Rose~\cite{rose1957:am},
  which is the complex conjugate of the $\mathcal{D}$ function defined by Bohr
  and Mottelson~\cite{bohr1998:v2}.}
\end{thebibliography}

\end{document}